\documentclass[aps,twocolumn, amsmath,amssymb, pra]{revtex4}
\usepackage{graphicx}
\usepackage{epsfig}

\begin{document}

\title{Fermion- and Spin-Counting in Strongly Correlated Systems}

\author{Sibylle Braungardt\(^1\), Aditi Sen(De)\(^1\), Ujjwal Sen\(^1\), Roy J. Glauber\(^2\), and Maciej
Lewenstein\(^{*,1}\)}
\affiliation{\(^1\)ICFO-Institut de Ci\`encies Fot\`oniques, Mediterranean Technology Park,
08860 Castelldefels (Barcelona), Spain\\
\(^2\)Lyman Laboratory, Physics Department, Harvard University, 02138 Cambridge, MA, U.S.A. \\
\(^*\)ICREA � Instituci\`o Catala de Ricerca i Estudis Avan{\c c}ats, 08010 Barcelona, Spain}

\begin{abstract}
%\vspace{.1cm}
We apply the atom counting theory
to strongly
correlated Fermi systems and spin models, which can be realized
with ultracold atoms. The counting distributions are typically
sub-Poissonian and remain smooth at quantum phase transitions, but
their moments exhibit critical behavior, and characterize quantum
statistical properties of the system. Moreover, more detailed
characterizations are obtained with experimentally feasible
spatially resolved counting distributions.
%\vspace{.1cm}
\end{abstract}

\maketitle

\def\com#1{{\tt [\hskip.5cm #1 \hskip.5cm ]}}

\section{Introduction}

\subsection{Particle- and spin-counting}

 Particle-wave duality is one of the most spectacular, and at the same time intriguing phenomena of quantum
 mechanics.
 Nevertheless, careful counting of particles, such as photons, in a given quantum mechanical state allows
 to fully reconstruct the wave nature and coherence properties of the state. The formulation of photon-counting
 theory in the frame of quantum electrodynamics 
 %by Glauber 
 \cite{roy} gave birth to modern quantum optics.
 Recent progress  in physics of ultracold atoms made possible to develop and apply  techniques of single
 atom counting to various systems. Since the pioneering experiments of
 Shimizu \cite{shimizu}, spectacular measurements of Hanbury Brown - Twiss effect for bosons \cite{allain},
 and fermions \cite{Aspectexperiment} have been performed with ultra-cold meta-stable Helium atoms.
 Esslinger's
 group employed cavity quantum electrodynamics  techniques to measure the pair correlation function in an atom laser   
 beam outgoing
 from a trapped Bose condensate \cite{essli}. These new detection methods allow in principle to
 measure full atom-counting distributions with spatial resolution (by counting
 only atoms in  a certain spatial region),
 and provide novel efficient ways of detection  of strongly correlated systems \cite{MaciejLewenstein}.

 Equally spectacular progress has been achieved in spin-counting, or in other words,
 measurements of total atomic spin for atoms with spin, or pseudo-spin degree of freedom.
 The idea of  quantum non-demolition  polarization spectroscopy (QNDPS), has been demonstrated
 in Ref.~\cite{Sorensen}.
 It employs the quantum Faraday effect: polarized light beam passed through the atomic sample, undergoes polarization
 rotation. Atomic fluctuations leave an imprint on the quantum fluctuations of the light, and vice versa.
 This idea  was recently extended to ultra-cold spinor gases \cite{ourPRL}, where it can be used to detect,
 manipulate, and even engineer various states of such systems. Amazingly, this method  allows also for a
 spatial resolution (when standing laser beams are employed) \cite{tobenat}.

\subsection{Main results}

In this paper, we show how the atom counting techniques can be
used to detect properties of strongly correlated systems. We
concentrate, in particular, on the case of fermion and/or spin
counting in one-dimensional (1D) optical lattices, that are equivalent, via
Jordan-Wigner transformation \cite{Sachdev}, to 1D spin chains. The
problem of spin counting for a local block of spins in the 1D
Ising model in a transverse field has been considered in a
beautiful work of Demler's group \cite{Demler}. Our paper is in a
sense complementary to Ref. \cite{Demler}.  First, we consider not
only on the Ising model, but on the whole family of asymmetric XY models,
characterized by the asymmetry parameter $\gamma$, in the
transverse field $h$. Second, employing ideas of Ref.
\cite{tobenat}, we calculate not only the counting distribution
for the total fermion number (total $Z$-spin component), but also
for ``effective" number, corresponding to certain spatial Fourier
components of the fermion density. While  for the considered
family of models, counting distributions are always smooth, their
cumulants  exhibit critical behavior, evident even for small
detection efficiencies.  The distributions are always
sub-Poissonian, but the sub-Poissonian character changes, as we
sweep $h$ from 0 to $\infty$. For small (large) $\gamma$, the
$h=0$ ($h=\infty$) distribution is always the narrower (broader) one.
Through the paper,  we use an elegant generalization of 
%Glauber's
the 
photon-counting theory to fermions, derived by Cahill and Glauber
within the formalism of Grassmann variables
\cite{Glauber+Cahill}. For the cases we consider, we obtain analytic
expressions for the counting distribution in terms of simple
recursion relations.

The paper is organized as follows. 
%We 
%introduce the model 
In Sec. \ref{sec_model}, we briefly describe the models of the 1D optical lattice that we consider, and the Jordan-Wigner 
transformation that can be used to diagonalize them. 
In the next section (Sec. \ref{sec_counting}), we derive the counting statistics of fermions in the systems described by
 these models; in particular, we discuss them  
%Under this section (Sec. \ref{sec_counting}), we discuss counting statistics 
for the Ising model (Subsec. \ref{sec_ising}), and more 
generally for the asymmetric XY model 
(Subsec. \ref{sec_XY}). In Subsec. \ref{sec_mean_variance}, we consider the means and variances of the counting distributions:
We derive recurrence relations that allow for easy calculation of these moments for an arbitrary number of particles in the 
system. 
We discuss also the generalization of our method to the case of Fourier components of the total spin in Subsec. \ref{sec_ditiyo}. 
Finally, we summarize our results in Sec. \ref{sec_obosesh}.

%\vspace{1cm}

\section{Fermi gas in an 1D optical lattice}
\label{sec_model}

\subsection{1D Fermi gases}

Let us  consider a family of models describing an one-dimensional Fermi gas in an optical lattice, described by the Hamiltonian
\begin{equation}
H=-\frac{J}{2}\sum_{j=0}^{N-1}\left[c_j^\dag c_{j+1}+\gamma c_j^\dag
c^\dag_{j+1}+\mbox{h.c.}-2gc_j^\dag c_j\right]+\frac{1}{2}Ng,\label{eq-ham1}
\end{equation}
where $J/2$ is the energy associated to fermion tunneling,
$g=h/J$, and $N$ is the number of sites. One way to realize such
Hamiltonian with ultracold atoms is to use a  Fermi-Bose mixture
in the strong coupling limit. In this limit,  the low energy
physics is well
 described by fermionic composites theory \cite{fbml}, in which fermions form composite
 objects with $0,1,\ldots$ bosons, or
bosonic holes repectively. The fermionic composites undergo
tunneling and interact via nearest neighbor interactions, which
may be repulsive  or attractive, weak or strong, depending on the
original parameters of the system, such as scattering lengths, etc.
In the case of weak  attractive interactions, the system undergoes,
at zero temperature, a transition into a ``$p$-wave'' superfluid, described
well by the Bardeen-Cooper-Schrieffer (BCS) theory, corresponding
exactly to the Hamiltonian (\ref{eq-ham1}).

\subsection{1D spin chains}

Using Jordan-Wigner transformation \cite{Sachdev}, one can
transform the Hamiltonian (\ref{eq-ham1}) into the one of a 1D
asymmetric XY spin chain in the transverse magnetic field $h$,
 \begin{equation}
H_{xy}=J\sum_{j=0}^{N-1}\left[(1+\gamma)S_j^xS_{j+1}^x+(1-\gamma)S_j^yS_{j+1}^y-
%\left[
\frac{h}{J}
%\right]
S_j^z\right], \label{eq-xy}
\end{equation}
where  $S_j^\alpha=\frac{1}{2}\sigma_j^\alpha$ are the spin $1/2$
operators at site $j$, proportional to Pauli matrices. The special
cases $\gamma=0$ (i.e. the so called symmetric XY, or XX limit)
and $\gamma =\pm 1$ can be realized with single species bosons in
the hard core (i.e. strongly repulsive) bosons limit
\cite{Sachdev,review}, or in a chain of double well sites filled
with bosons interacting via weak dipolar forces \cite{jaksch},
respectively. In general, one should use a two component Bose-Bose
and Fermi-Fermi  mixture, which, in the strong coupling limit, and
in the Mott insulator state with one atom per site, is described by
an asymmetric (XXZ) Heisenberg model (cf. \cite{laurent}) in the Z
oriented field. By appropriate tuning of the scattering lengths
via Feshbach resonances, one can set the $S_{j+1}^z S_j^z$
coupling to zero, i.e. achieve the XX model in the transverse
field. In order to introduce the asymmetry $\gamma$, one should
additionally introduce tunneling assisted with a laser or
microwave induced double spin flip. For this aim, one should make
use of the resonance between on-site two atom ``up-up'' and
``down-down'' states, without disturbing ``up-down''
configurations.

\subsection{Jordan-Wigner transformation}

As it is well known, Jordan-Wigner transformation works for open
chains, and in particular for an infinite chain. We will
nevertheless assume periodic boundary conditions to solve the
fermion model  (\ref{eq-ham1}) using  Fourier and Bogoliubov
transformations (see e.g. \cite{eita-McCoy}). For large $N$, such
precedure gives the right leading behaviour. We define Fourier
transformed operators as 
\begin{equation} c_j^\dag =
\sum_{k=0}^{N-1}\exp(-ij\Phi_k)a_k^\dag,
\end{equation} and  
\begin{equation}
c_j =
\sum_{k=0}^{N-1}\exp(ij\Phi_k)a_k,
\end{equation} 
 where $\Phi_k=2\pi k/N$. We
perform then the Bogoliubov transforms 
\begin{equation}
a_k=u_kd_k-iv_kd_{N-k}^\dag, \quad
a_k^{\dag}=u_kd_k^\dag+iv_kd_{N-k},
\end{equation}
 where $u_k$, $v_k$ are real
numbers satisfying 
\begin{eqnarray}
&&u_k^2+v_k^2=1, \nonumber \\
&&u_{N-k}=u_k  \quad \mbox{and} \quad
v_{N-k}=-v_k,
\end{eqnarray} so that we can write 
\begin{equation}
u_k=\cos{\frac{\theta}{2}}, \mbox{ and } v_k=\sin{\frac{\theta}{2}}.
\end{equation}
When  \begin{equation}\tan{\theta}=\frac{\gamma\sin{\Phi_k}}{\cos{\Phi_k}-g},\end{equation}
the Hamiltonian reduces then to the noninteracting fermions
Hamiltonian,
\begin{equation}
H=\frac{1}{2}\sum_{k=0}^{N-1}\epsilon_kd_k^\dag d_k,
\end{equation}
with
\begin{equation}
\epsilon_k=2\sqrt{(\cos{\Phi_k}-g)^2+\gamma^2\sin^2{\Phi_k}}.
\end{equation}
The ground state is thus the vacuum of the $d_k$ operators. For
$\gamma>0$ the spectrum is everywhere gapped, except at the
critical point $g_c=1$. For $\gamma =0$, $d_k$'s coincide with
$a_k$'s or $a_k^\dag$'s, and the ground state is a Fermi sea. For
$-1\le g\le 1$ the spectrum is then gapless and the system
critical. Note that the number of original fermions $\hat
N_f=\sum_{i=0}^{N-1} c_i^\dag c_i$, as well as the the total
Z-component of the spin, $\hat S^z=\sum_{i=0}^{N-1} S_i^z=\hat N_f
-1/2$ are not conserved, except at $\gamma =0$.

\section{Fermion Counting Statistics}
\label{sec_counting}

\subsection{Fermion counting distributions}

Let us now turn to counting procedures. For the case of fermions, one should think about the analogue approach
as one used in the experiments on
metastable Helium. For spins, one could use directly QNDPS to measure the distribution of $\hat S_z$, or even its
spatially resolved version \cite{tobenat}. An alternative way would be to switch
off the Hamiltonian (\ref{eq-xy}) (by switching off lasers), and induce spontaneous Raman transition from the state ``up'' to some side level.
Counting of spontaneously emitted photons would correspond then to counting of ``up'' spins''

Mathematically, 
%as shown by Glauber 
as  known 
for photons \cite{roy}, and generalized by Cahill and Glauber for fermions \cite{Glauber+Cahill},
the probability of detecting $m$ photons in a given interval of
time can be expressed as the $m$th derivative with respect to a
parameter $\lambda$ of the generating function
$\mathcal{Q}(\lambda)$ as
\begin{equation}\label{eq-p}
p(m)=\frac{(-1)^m}{m!}\frac{d^m}{d\lambda^m}\mathcal{Q}\Big|_{\lambda=1},
\end{equation}
where $\mathcal{Q}(\lambda)$  is the expectation value of a
normally ordered exponential
$\mathcal{Q}(\lambda)=\mbox{Tr}(\rho:e^{-\lambda\mathcal{I}}:)$.
The operator $\mathcal{I}$ is a space-time integral of the product
of the positive-frequency and negative-frequency parts of the
quantum fields describing particles to be counted.
The mean values of normally ordered products can be calculated in a particularly convenient 
and elegant way using the Grassmann variables formalism, introduced in \cite{Glauber+Cahill}.
In the case of counting the total number of particles, we have
$ \mathcal{I}=\kappa\sum_{j=0}^{N-1} c_j^\dag
c_j=\kappa\sum_{j=0}^{N-1} \sigma_j^\dag\sigma_j=
\kappa\sum_{k=0}^{N-1} a_k^\dag a_k$ ,  where
$\kappa=1-\exp({-\zeta t})$, while $0\le \zeta\le 1$ is the
detector efficiency, and $t$ is the exposure time. For the
spatially resolved QNDPS, $\mathcal{I}=\kappa\sum_{j=0}^{N-1}
\sigma_j^\dag\sigma_j\cos({\bf k}_L{\bf r}_j)$, where ${\bf k}_L$
is the wave vector of the standing wave used for detection, and
${\bf r}_j$ is the  position  of the $j$-th site.

%%%%%%%%%%%%%%%%%%%%%

For counting the total number of particles, the generating function $\mathcal{Q(\lambda)}$ can be written
as
\begin{equation}
\mathcal{Q}(\lambda)=\mbox{Tr}(\rho:e^{-\lambda\kappa\sum_{k=0}^{N-1} a_k^\dag
a_k}:).
\end{equation}
The operators $a^\dag_ka_k$ commute for different $k$, so that the
expression for $\mathcal{Q}$ can be rewritten as
\begin{eqnarray}
\mathcal{Q}(\lambda)&=&\mbox{Tr}(\rho
:\prod_{k=0}^{N-1}(e^{-\lambda\kappa
a_k^\dag a_k}):) \nonumber \\
&=&\mbox{Tr}(\rho :\prod_{k=0}^{N-1}(1-\lambda\kappa a_k^\dag
a_k+\lambda^2\kappa^2a_k^\dag a_ka_k^\dag a_k+...):) \nonumber \\
&=&\mbox{Tr}(\rho \prod_{k=0}^{N-1}(1-\lambda\kappa a_k^\dag
a_k))\nonumber
\\&=&\mbox{Tr}(\rho
\prod_{k=1}^{N/2}(1-\lambda\kappa a_k^\dag a_k)(1-\lambda\kappa
a_{N-k}^\dag a_{N-k})),\nonumber
\end{eqnarray}
as $:a_k^\dag a_ka_k^\dag a_k:=a_k^\dag a_k^\dag a_ka_k=0$, etc.

The terms $a^\dag_ka_k$ and $a^\dag_{N-k}a_{N-k}$ can then be
expressed in terms of the $d$ fermions:
\begin{eqnarray}
a^\dag_ka_k=(u_kd_k^\dag+iv_kd_{N-k})(u_kd_k-iv_kd_{N-k}^\dag), \nonumber\\
a_{N-k}^\dag
a_{N-k}=(u_kd_{N-k}^\dag-iv_kd_{k})(u_kd_{N-k}+iv_kd_{k}^\dag). \nonumber
\end{eqnarray}

\subsection{Generating function for the ground state}

We consider 
the counting statistics of the $c$ fermions in
the ground state of the Hamiltonian, i.e. \emph{in the vacuum state of
 $d$ fermions}. 

The trace in the generating function can be now easily  calculated by the formalism of Grassmann variables 
%$P$ representation for the 
%density operator $\rho$ 
%introduced in 
\cite{Glauber+Cahill}.
The $P$ representation for the 
density operator $\rho$ is
\begin{equation}
\rho=\int d^2\vec{\alpha} P(\vec{\alpha})|\vec{\alpha}\rangle \langle \vec{\alpha}|,
\end{equation}
where $|\vec{\alpha}\rangle$ are the fermionic coherent states, as defined in \cite{Glauber+Cahill}.
Using the $P$ representation, the mean values of normally ordered products of $d$-fermions can then be calculated as
\begin{eqnarray}
\mbox{Tr}(\rho d_k^{\dag n}d_l^m)=\int d^2\vec{\alpha} P( \vec{\alpha} )\langle\vec{\alpha}|d_k^{\dag n}d_l^m|\vec{\alpha}\rangle\nonumber \\
=\int d^2\vec{\alpha} P(\vec{\alpha})\alpha_k^{*n}\alpha_l^m, \label{eq-trace}
\end{eqnarray}
where the $\alpha_i$ are Grassmann variables, and are defined by the eigen-equation \(d|\alpha_i\rangle = \alpha_i |\alpha_i\rangle\).
For the vacuum state of the $d$-fermions,
\begin{equation}
\rho=|{0...0}\rangle \langle{0...0}|,
\end{equation}
the $P$-function is given by
\begin{equation}
P(\alpha)= \int d^2\vec{\xi} \exp\left(\sum_i(\alpha_i\xi_i^*-\xi_i\alpha_i^*)\right)=\delta(\vec{\alpha}). \label{eq-P}
\end{equation}
Evaluating Eq. (\ref{eq-trace}) using Eq. (\ref{eq-P}), we get the relations
\begin{equation}
\mbox{Tr}(\rho d_k^{\dag n}d_l^m)=\int d^2 \vec{\alpha} \prod_i(\alpha_i^*)^{n_i}\alpha_i^{m_i}\delta(\vec{\alpha})=0, \label{normalordering}
\end{equation}
and
\begin{equation}
\label{qwert}
\mbox{Tr}(\rho)=\int d^2\vec{\alpha}\delta(\vec{\alpha})=1. 
\end{equation}
The relevant remaining terms in the product
$(1-\lambda\kappa a_k^\dag a_k)(1-\lambda\kappa a_{N-k}^\dag
a_{N-k})$ in the generating function are thus 
%given by
\begin{eqnarray}
1-\lambda\kappa v_k^2d_{N-k}d_{N-k}^{\dag}-\lambda\kappa
v_k^2d_kd_k^\dag\nonumber\\
+\lambda^2\kappa^2v_k^4d_{N-k}d_{N-k}^\dag
d_{k}d_{k}^\dag-\lambda^2\kappa^2v_k^2u_k^2d_{N-k}d_kd_{N-k}^\dag
d_k^\dag\nonumber 
%= 1-\lambda\kappa v_k^2d_{N-k}d_{N-k}^{\dag}-\lambda\kappa
%v_k^2d_kd_k^\dag\nonumber
%\\
%+\lambda^2\kappa^2v_k^4d_{N-k}d_{N-k}^\dag
%d_{k}d_{k}^\dag+\lambda^2\kappa^2v_k^2u_k^2d_{N-k}d_{N-k}^\dag d_k
%d_k^\dag \nonumber 
\end{eqnarray}
Elementary calculations using the relations (\ref{normalordering}) and (\ref{qwert}) yield
\begin{equation}
\mathcal{Q}(\lambda)=\prod_{k=1}^{N/2}\Big(1-2\lambda\kappa
v_k^2+\lambda^2\kappa^2v_k^2\Big). \label{eq-Q}
\end{equation}
 At this point it is convenient to
introduce the distribution function $p(m,M)$ of counting $m$
particles for $M$ pairs of modes. It is   given by the same
expression as before, but with the product in Eq. (\ref{eq-Q})
limited to $M/2$ terms.

%%%%%%%%%%%%%%%%%%%%%%

\subsection{Counting statistics}

The counting distribution is calculated from the generating
function by the relation in Eq. (\ref{eq-p}).
%,
%\begin{equation}
%p(m)=\frac{(-1)^n}{n!}\frac{d^n}{d\lambda^n}\mathcal{Q}|_{\lambda=1}.
%\end{equation}
We use the generalized Leibniz rule,
\begin{eqnarray}
&&\frac{d^m}{d\lambda^m}\prod_{k=1}^{N}f_k(\lambda) \nonumber \\
&=& \sum_{n_1+...+n_N=n}\left(n
\atop
n_1,n_2,...,n_N\right)\prod_{k=1}^N\frac{d^{n_k}}{d\lambda^{n_k}}f_k(\lambda), \nonumber
\end{eqnarray}
where the generalized Newton's symbol is given by
\begin{equation}
\left(n \atop n_1,n_2,...,n_N\right)=\frac{n!}{n_1!n_2!...n_N!}, \nonumber
\end{equation}
%\vspace{1cm}
%end{equation}
to derive a recurrence relation, to calculate the distribution for
$(M+1)$ modes, given the distribution for $M$ modes.

The distribution function $p(m,M)$ for $M$ modes is given by
\begin{eqnarray}\label{gambat}
&&p(m,M)=\frac{(-1)^m}{m!}\frac{d^m}{d\lambda^m}\mathcal{Q}\Big|_{\lambda=1}\nonumber\\
&&=\frac{(-1)^m}{m!}\sum
%_{{l_1,...l_N\atop {l_1+...+l_{N}=m \atop
%l_j=0,1}}}
\frac{m!}{l_1!l_2!...l_N!}\prod_{j=1}^{M/2}\frac{d^{l_j}}{d\lambda^{l_j}}(1+A\lambda+B\lambda^2),\nonumber\\
\end{eqnarray}
where the summations run over \(l_1, \dots, l_M\) such that \(l_1
+ \dots +l_M =m\), where \(l_j = 0\), \( 1\), or \( 2\), for
\(j=1,\dots,M\).

%The distribution for $M+1$ modes can then be calculated as
%follows.
%\begin{eqnarray}
%p(m,M+1)=(1+A_{M+1}+B_{M+1})p(m,M)\nonumber\\
%-(A_{M+1}+2B_{M+1})p(m-1,M)\nonumber\\
%+B_{M+1}p(m-2,M) \label{asol_jinis}
%\end{eqnarray}
%Therefore, starting from $p(0,1)=1-2\kappa
%v_1^2+\kappa^2(v_1^4+u_1^2v_1^2)$, $p(1,1)=2\kappa
%v_1^2+2\kappa(v_1^4+u_1^2v_1^2)$ and
%$p(1,2)=\kappa(v_1^4+u_1^2v_1^2)$, we can use the recurrence
%relation (\ref{asol_jinis}) to calculate the counting distribution
%for an arbitrary number of modes.

%%%%%%%%%%%%%%%%%%%%%%%

%Using  the generalized Leibnitz rule,
 We can now 
%obtain a formal expression for $
%p(m,M)$,  
%and 
derive the recursive relation
\begin{eqnarray}
p(m,M+1)&=&  \sum_{i=0}^{2} {\cal P}_ip(m-i,M)
%%+{\cal P}_1p(m-1,M)\nonumber\\
%%&+&{\cal P}_2p(m-2,M), 
\label{asol_jinis}
\end{eqnarray}
where 
\begin{eqnarray}
{\cal P}_0 &= & 1-2\kappa v_{M+1}^2+\kappa^2v_{M+1}^2, \nonumber \\
{\cal
P}_1 &=& 2\kappa v_{M+1}^2-2\kappa^2v_{M+1}^2, \nonumber \\
{\cal
P}_2 &=& 1-{\cal P}_0-{\cal P}_1
\end{eqnarray} 
are the probabilities of detecting
0,1, or 2 particles in the modes $M+1$ and $N-M-1$. Therefore,
starting from $p(0,1)=1-2\kappa v_{1}^2+\kappa^2v_{1}^2$,
\(p(1,1)= 4 \kappa v_1^2\) 
% 2\kappa v_1^2-2\kappa v_1^2$ 
and $p(2,1)=\kappa v_1^2$,
we can use the recurrence relation (\ref{asol_jinis}) to calculate
the counting distribution for an arbitrary number of modes.

Let us turn now to our results and discuss the counting statistics for different values of \(\gamma\).
In the figures that we plot below (except in Fig. \ref{fig-splitting} in Subsec. \ref{sec_split}), 
we choose a value of the total number of modes, \(N\), such that the 
corresponding quantities (distribution, mean, variance, etc.) have already converged. In the cases that we consider,
such convergence occurs for \(N \approx 300\).
%, and
%$N=5000$.

\subsection{Transverse Ising model}
\label{sec_ising}

\begin{figure}
\begin{center}
\epsfig{figure= 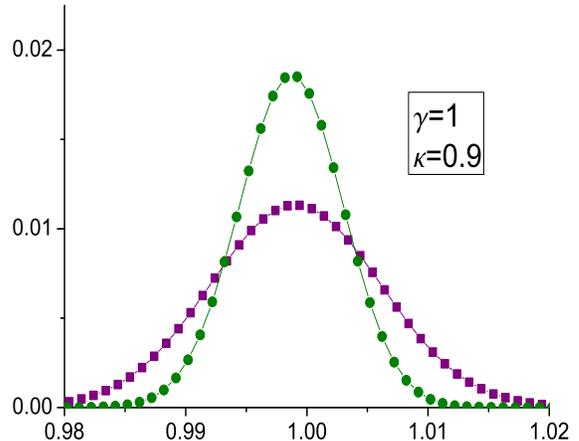, height=.3\textheight,width=0.47\textwidth}
\end{center}
\caption{Counting statistics
of the transverse Ising model. The horizontal axis is
(\(m-\bar{m})/N+1\). The vertical axis is of the corresponding
probability. The curve with purple squares is for $h/J= 0.01$,
while the one with green circles is for $h/J = 10$. The QPT of
this model is at \(h/J=1\). Both the distributions are
sub-Poissonian. However, the counting distribution becomes much
narrower in the case when $h/J
> 1$ than the situation when $h/J<1$. In this case, we have taken
the efficiency $\kappa$ as $0.9$.
% and N=5000.
} \label{comaparison}
\end{figure}

The counting distributions for the transverse Ising model
(transverse XY model with \(\gamma = 1\)) for two exemplary values
of the field parameter \(g=h/J\) are shown in Fig.
\ref{comaparison}. The Ising model has a quantum phase transtition
at \(g=1\) \cite{Sachdev}, and  one exemplary value of \(g\) is
chosen below the QPT, and the other above it. The difference in
behavior is clearly seen. ($\overline{m}$ and  $var$ denote the mean and variance of the distribution, respectively.) 
Below, it will be more clearly revealed
by looking at the mean and the variance of the distribution.

\subsection{Transverse XY model: ``Transition anisotropy''}
\label{sec_XY}

\begin{figure}
\begin{center}
\epsfig{figure= 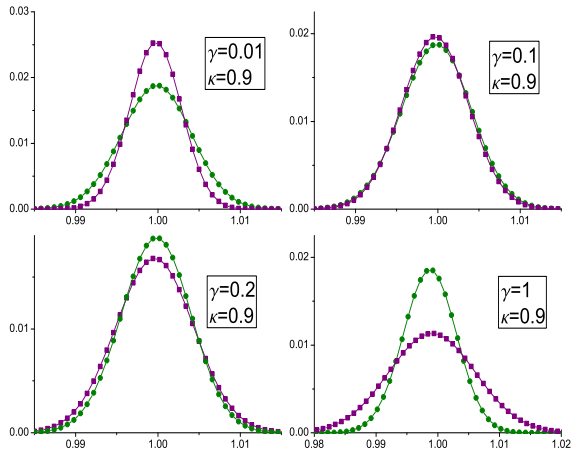, height=.3\textheight,width=0.47\textwidth}
\caption{Fermion counting distribution as a function of $(m-\overline{m})/N +1$ (horizontal axis) 
for $\kappa=0.9$,
%, \(N=5000\), 
and for the indicated values
of $\gamma$.  Purple squares   correspond to $h/J \to
0$, while  green circles to
$h/J \to \infty$. The transition anisotropy is here at \(\gamma \approx 0.1\).}\label{comaparison1}
\end{center}
\end{figure}

 In Fig. \ref{comaparison1}, we plot counting distributions as a function of $(m-\overline{m})/N+1$ for four
values of \(\gamma\),  for a fixed value of the  efficiency  \(\kappa = 0.9\), and  for two extreme values of \(g\):
\(g\to 0\) and  \(g \to \infty\). 
  Note that all the distributions presented in Fig. \ref{comaparison1} are smooth
and their widths ($\simeq\sqrt{var}/N$) are of order of 0.01.
Since, as we argue below, $\overline{m}\simeq \kappa N$, all the 
distributions are  sub- Poissonian, because $var\le \bar m$,
despite the finite detection efficiency. For \(\gamma \to 0\), the
distribution for \(g\to 0\) is narrower than that for \(g \to
\infty\). This tendency is inverted in the Ising
 model, when the distribution for \(g\to 0\)
has a \emph{larger} variance than the one for \(g\to \infty\).
At, what we call,   \emph{transition anisotropy}  \(\gamma \approx 0.1\),
the distributions for \(g \to 0\) and \(g \to
\infty\) practically coincide.

This transition anisotropy depends on the
efficiency \(\kappa\), and it moves to \(\gamma \to 0\), as
\(\kappa \to 1\). This indicates that the probability distribution
of counting can distinguish the two universality classes (the XX, with \(\gamma =0\), and
the Ising, with \(\gamma >0\)) among the XY models on a chain. In the limit of \(\kappa
\to 1-\), only the model with \(\gamma \to 0\) has lower variance
for \(g\to 0\) as compared to \(g \to \infty\), while all the
other XY models (with \(\gamma \ne 0\)) have the opposite behavior.
%................................................bhabtey hobey!!!

\subsection{Recurrence relations for mean and variance}
\label{sec_mean_variance}

\begin{figure}[tbp]
\begin{center}
\epsfig{figure=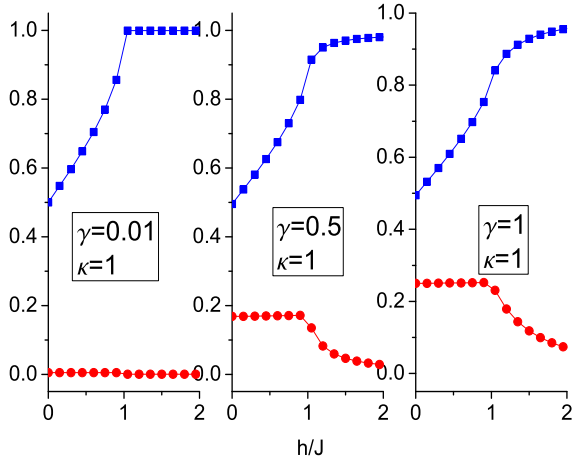,
height=0.3\textheight,width=0.47\textwidth}
\caption{Mean $\overline{m}/N$ (blue squares) and variance $var/N$ (red circles)  of the
fermion counting distribution as a function of \(g=h/J\) for $\kappa=1$, and indicated values of $\gamma$.} \label{fig-gammas}
\end{center}
\end{figure}

\begin{figure}[tbp]
\begin{center}
\epsfig{figure=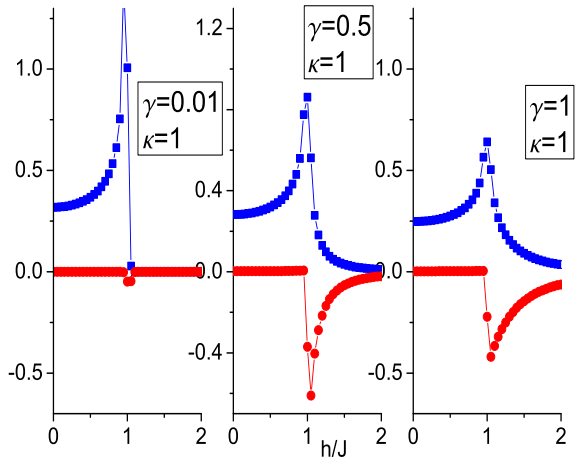,
height=.3\textheight,width=0.47\textwidth}
\caption{The derivatives of the  means and variances are plotted against the transverse field \(h/J\) (horizontal axis),
 for \(\gamma =0.01\), \(\gamma =0.5\),
and \(\gamma = 1\).
%Therefore the extreme left is for the XX model, while the extreme right is for the Ising.
Blue  squares denote the derivatives of the means, while
red circles denote the derivatives of the variances, in the
respective cases. Also, $\kappa=1$. The QPTs of all the models at
\(g=1\) are clearly visible. } \label{fig-gammasderivatives}
\end{center}
\end{figure}

\begin{figure}
\begin{center}
\epsfig{figure= 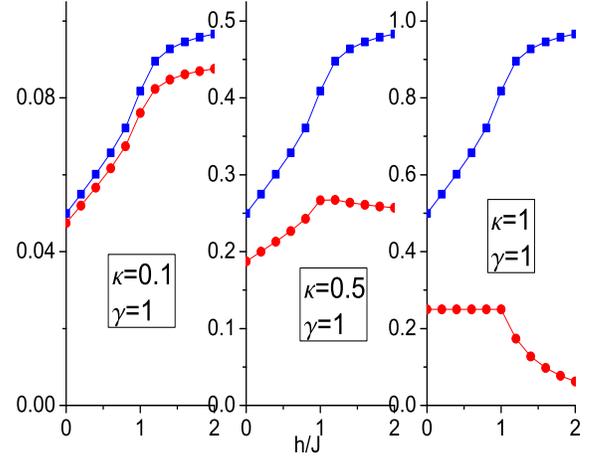,
height=.3\textheight,width=0.47\textwidth}
\caption{Mean $\overline{m}/N$ (blue squares) and variance $var/N$ (red circles)  of the
fermion counting distribution as a function of \(g=h/J\) for $\gamma=1$, and indicated values of $\kappa$.}
\label{fig-means-variances}
\end{center}
\end{figure}

In order to understand the properties of counting distributions
better, we look at the mean and variance, which can be calculated
from the following recurrences:
\begin{eqnarray}
\overline{m_{M+1}}=\overline{m_M}+2\kappa v_{M+1}^2,\\
%%\end{equation}
%%and
%\begin{eqnarray}
var_{M+1}=\overline{m_{M+1}^2}-\overline{m_{M+1}}^2 \nonumber\\
=var_{M}+4\kappa^2v_{M+1}^2(1-v_{M+1}^2)
\end{eqnarray}
Since $\overline{m_1}$ and $var_1$ can be  trivially calculated,
the mean and variance can be obtained by these relations for an
arbitrary number of modes. The recurrences imply that the mean
$\overline{m_{N}}\le \kappa N$; we find typical value of
$\overline{m_{N}}$  indeed of order of $\kappa N$. On the other hand, the variance
%on the other hand, 
$var_N\le \kappa^2N$. Both quantities show
singular behavior in the thermodynamical limit at criticality. In
particular, for the transverse Ising model ($\gamma=1$), near the
critical point \(g=g_c \equiv 1\), the mean \(\overline{m}\) can
be written in terms of  elliptic integrals of first and second
kind, and  can be expressed as \cite{ghyama2}
%behaves as
\[
\overline{m} \approx - \frac{1}{2\pi} (g-g_c) \ln \left|g-g_c\right| - \frac{1}{\pi},
\]
so that 
\[ d\overline{m}/{dg} \approx - (\ln |g-g_c|+1)/{2\pi}.\]
Since all models with $\gamma \ne 0$ belong to the same
universality class, they all present the same singular behavior
\cite{Sachdev}. This is contrasted with the case of XX model,
which belongs to a different universality class. The singular
behavior is clearly seen in the plots of $\overline{m}/N$ and
$var/N$ obtained for finite $N\simeq 300$ and ideal $\kappa=1$
(see Fig. \ref{fig-gammas}). For finite values of $\gamma$, the
variance shows a jump in the first derivative, while the first
derivative of the mean tends to ``infinity" at $g_c$. This behavior
is better seen, when one plots directly the derivatives of
$\overline{m}$ and $var$ (see Fig. \ref{fig-gammasderivatives}). 
This behavior changes drastically as
$\gamma\to 0$.  The variance tends then to zero (in the symmetric
XX model the particle number is conserved), and the mean has a
diverging derivative for $g<g_c$, and is constant for $g>g_c$.
Amazingly, although finite detector efficiency obviously smoothes
out the curves, the signatures of the singularities are clearly
visible even for $\kappa =0.5$ (see Fig.
\ref{fig-means-variances}). A clear change of behavior of the
curves is visible even at $\kappa=0.1$!  Note, that in all
considered cases so far, the variance $var/N<\overline{m}/N$, i.e.
all distributions are sub-Poissonian. Note, however, that going
from anti-ferromagnetic to the \emph{ferromagnetic} case, does not affect the
variance, but replaces $\overline{m}/N\to (1/2-\overline{m}/N)$.
In that case we do observe a transition from sub-Poissonian
behavior at small $g<g_t$, to (weakly) super-Poissonian for
$g>g_t$; the value of $g_t$ tend to $g_c$ from below as $\gamma\to
0$.

%%%%%%%%%%%%%%%%%%%%%%%%%%%%%%%%%%%%%%%%%%%%%%%%%%
%%%%%%%%%%%%%%%%%%%%%%%%%%%%%%%%%%%%%%%%%%%%%%%%%%%

%%%%%%%%%%%%%%%%%%%%%%%%%%%%%%%%%%%%%%%%%%%%%%%%%%
%%%%%%%%%%%%%%%%%%%%%%%%%%%%%%%%%%%%%%%%%%%%%%%%%%%

\subsection{Even versus odd splitting}
\label{sec_split}

\begin{figure}[tbp]
\begin{center}
\epsfig{figure=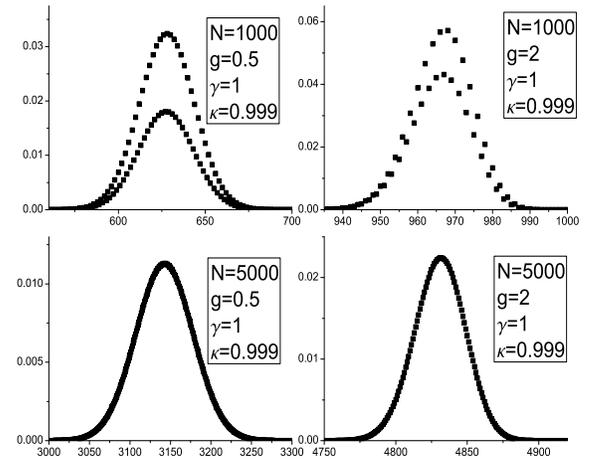,
height=.3\textheight,width=0.47\textwidth}
\caption{Even versus odd splitting for $\kappa=0.999$ in the Ising
model. For N=1000 the probability distribution splits up, whereas
for N=4000 there is no splitting.}\label{fig-splitting}
\end{center}
\end{figure}

The Bogoliubov transformation used to solve the considered models
can be regarded  as a ``squeezing'' or ``pairing'' transformation. The
ground state that we investigated is  analogous to BCS states of
semiconductors, i.e. they involve fermion (Cooper-like) pairs.
Thus, in the ideal case of $\kappa=1$, the counting distributions are
exactly zero for odd numbers of particles. In practice, for finite
values of $N$ and $\kappa<1$, the distributions oscillate between
larger values for even, and small for odd number of counts. This
behavior is very strongly affected by $\kappa<1$, since at  finite
efficiency,  one may easily miss single atoms from the Cooper
pairs, and obtain   odd counts. In effect, for a given value of
$N$, the even-odd asymmetry is visible only for $\kappa$ close
enough to 1. Similarly, the even-odd asymmetry is strongly
affected by the finite size effects - for a given value
of $\kappa<1$ it is visible only for $N$ small enough
 (see Fig. \ref{fig-splitting}).

\subsection{Counting spatial Fourier components of the fermion density}
\label{sec_ditiyo}

Finally, let us point out that the methods proposed in \cite{tobenat} allow for measurements
 of various kinds of Fourier components of the total spin; in terms of particle counting, these methods
  allow  for instance to count particles in every second, every third site, etc. Our theory is easily
  generalized to such situations. 

In the case when we count  every second $c$
  fermion,
we have to express $b_{2j}^\dag b_{2j}=c_{2j}^\dag c_{2j}$ in
terms of the $d$ fermions. 
%%%%%%%%%%%%%%%%%%%%%%%%%
As before, as a first step we do the
Fourier transform:
%as in Eq. (\ref{eq-fourier}):
\begin{eqnarray}
c_{2j}^\dag &=& \sum_{k=0}^{N-1}\exp(-2ij\Phi_k)a_k^\dag, \nonumber \\
%\,\,\,\,\,\,
c_{2j}&=&\sum_{k=0}^{N-1}\exp(+2ij\Phi_k)a_k.
\end{eqnarray}
The expression $\sum_{j=0}^{N/2-1}c_{2j}^\dag
c_{2j}=\frac{1}{2}\sum_{j=0}^{N-1}c_{2j}^\dag c_{2j}$ can thus be
%reformulated to
written as
\begin{equation}
\sum_{j=0}^{N/2-1}c_{2j}^\dag
c_{2j}=\frac{1}{2}\sum_{k,k'}\frac{1-\exp \left(4\pi
i(k-k')\right)}{1- \exp \left(4\pi i(k-k')/N\right)}a_k^\dag
a_{k'},
\end{equation}
which is non vanishing for $k-k'=0$ or $|k-k'|=\frac{N}{2}$.
Finally
\begin{eqnarray}
&&\sum_{j=0}^{N/2-1}c_{2j}^\dag c_{2j} \nonumber \\
&=&\frac{1}{2}\sum_{j=0}^{N/2-1}a_k^\dag
a_k+a^\dag_{k+N/2}a_{k+N/2}+a_k^\dag a_{k+N/2}+a_{k+N/2}a_k \nonumber \\
&=&\frac{1}{2}\sum_{j=0}^{N/2-1}(a_k^\dag+a_{k+N/2}^\dag)(a_k+a_{k+N/2}).
\end{eqnarray}
We can now calculate $\mathcal{Q}(\lambda)$ as follows:
%to get the
%distribution.
\begin{eqnarray}
\mathcal{Q}(\lambda)=\mbox{Tr}(\rho
:\prod_{k=0}^{N/2-1}e^{-\frac{1}{2}\lambda\kappa(a_k^\dag+a_{N/2+k}^\dag)(a_{k}+a_{N/2+k})}:)\nonumber\\
=\prod_{k=0}^{N/2-1}\Big(1-\frac{1}{2}\lambda\kappa(a_k^\dag+a_{N/2+k}^\dag)(a_{k}+a_{N/2+k})\Big)\nonumber\\
=\prod_{k=1}^{N/4}\Big((1-\frac{1}{2}\lambda\kappa(a_k^\dag+a_{N/2+k}^\dag)(a_{k}+a_{N/2+k}))\nonumber\\
\times (1-\frac{1}{2}\lambda\kappa(a_{N-k}^\dag+a_{N/2-k}^\dag)(a_{N-k}+a_{N/2-k}))\Big).
\end{eqnarray}

\begin{figure}
\begin{center}
\epsfig{figure= 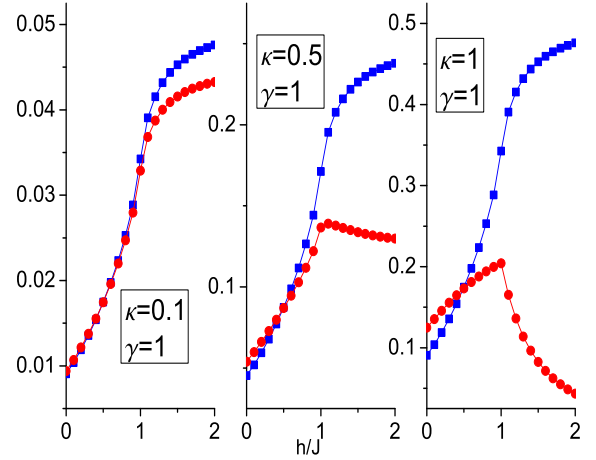,
height=.3\textheight,width=0.47\textwidth}
\caption{Mean $\overline{m}/N$ (blue squares) and variance $var/N$ (red circles)  of the
counting distribution of every second fermion as a function of \(g=h/J\) for $\gamma=1$, and indicated values of $\kappa$.} 
\label{fig-means-variances-sec}
\end{center}
\end{figure}

After performing the Bogoliubov transform,
% as in Eq. (\ref{eq-bog})
and keeping the relevant terms for the vacuum state of the \(d\)
fermions, the generating function $\mathcal{Q}$  is given by
\begin{equation}
\mathcal{Q}(\lambda)=\prod_{k=1}^{N/4}\Big(1-2\lambda\kappa
v_k^2+\lambda^2\kappa^2 v_k^2 \Big),
\end{equation}
%\subsection{Mean and Variance when counting every second spin}
which is in the same form
as in Eq. (\ref{eq-Q}), with the product restricted, however, to
$N/4-1$ terms.
We then  easily derive analogous recurrences
as in the cases considered so far.
Fig. \ref{fig-means-variances-sec} show the behavior of the mean
and the variance, when counting every second spin, in the
transverse Ising model. Note that the traces of singular behavior at $g=g_c$
persist. What is perhaps more interesting is that the general behavior is more rich. 
In particular, there is a crossing from sub- to
super-possoinian behavior at $g=0.5$. For \(\gamma \to 0\) the
point of crossing moves to zero, and the variance disappears.

\section{Summary}
\label{sec_obosesh}

Summarizing, we have formulated and applied fermion and spin
counting theory
%, 
%\'a la Glauber, 
to a family of one-dimensional strongly
correlated systems that can be realized and detected  with
ultracold atoms. The counting distributions exhibit traces of
singularities at criticality, that persist even at low detection
efficiencies. They show various kinds of rich behavior, such as
transitions from sub- to super-Poissonian character and even-odd
oscillations. 

\acknowledgments

We acknowledge support from the Spanish MEC
(FIS-2005-04627, Consolider Ingenio 2010 QOIT, Acciones Integradas, \& Ram{\'o}n y
Cajal), ESF Programmes QUDEDIS and Euroquam FERMIX, DAAD (German Academic Exchange Service), 
the Ministry of Education of the Generalitat de Catalunya, and EU IP
SCALA.

\end{document}